# The Automatic Creation of Concept Maps from Documents Written Using Morphologically Rich Languages

Krunoslav Zubrinic[1], Damir Kalpic[2], Mario Milicevic[1]

[1]{krunoslav.zubrinic, mario.milicevic}@unidu.hr
Department of Electrical Engineering and Computing, University of Dubrovnik, Cira Carica 4, Dubrovnik, Croatia

[2]damir.kalpic@fer.hr
Faculty of Electrical Engineering and Computing, University of Zagreb, Unska 3, Zagreb, Croatia

**Corresponding author**
Krunoslav Zubrinic
Department of Electrical Engineering and Computing, University of Dubrovnik, Cira Carica 4, HR-10000 Dubrovnik, Croatia
E-mail: krunoslav.zubrinic@unidu.hr
Phone: +385 20 445 742
Fax: +385 20 445 770

## Abstract

A concept map is a graphical tool for representing knowledge. They have been used in many different areas, including education, knowledge management, business and intelligence. Manually constructing concept maps can be a complex task; the unskilled person may encounter difficulties in determining and positioning concepts relevant to the problem area. An application that recommends concept candidates and their position in a concept map can significantly help the user in that situation. This paper gives an overview of different approaches to automatic and semi-automatic creation of concept maps from textual and non-textual sources. The concept map mining process is defined, and one method suitable for the creation of concept maps from unstructured textual sources in highly inflected languages such as the Croatian language is described in detail. Proposed method uses statistical and data mining techniques enriched with linguistic tools. With minor adjustments, that method can also be used for concept map mining from textual sources in other morphologically rich languages.

**Highlights**

- A concept map mining process is defined.
- An overview of different approaches to automatic and semi-automatic creation of concept maps from textual and non-textual sources is given.
- One method suitable for the creation of concept maps from unstructured textual sources in highly inflected languages is proposed.
- The proposed method uses statistical and data mining techniques enriched with linguistic tools.

# 1   Introduction

A concept map (CM) is a graphical tool that has been successfully used for organizing and representing knowledge. It includes concepts, usually signified by nouns or noun phrases, and relationships between them indicated by a line linking two concepts. Labelling a line with a verb or a verb phrase creates a concept-label-concept chain that can be read as a sentence. This chain is called a proposition (Novak & Cañas 2008).

There has been a remarkable growth in the use of CMs throughout the world over the past decade. The most prevalent applications of concept mapping are facilitating meaningful learning, and capturing and archiving expert knowledge in a form that would be easy to use by others. Furthermore, CMs have been known to be an effective tool to organize and navigate through large volumes of information.

In personal learning, a CM can be used as a tool that represents a learning plan, which consists of a set of goals that a person hopes to achieve within a specific period. For most learners it is difficult to begin with a "blank sheet" and start to build a map for a chosen topic of interest. A skeleton map provided by an expert can make it easier for the learner to start that process. In personal learning, it is difficult to find experts for specific learning fields. Therefore, an information system that behaves like an expert and provides the skeleton of a CM can be very helpful in such situations.

The automatic or semi-automatic creation of CMs from documents is called concept map mining (CMM) (Villalon & Calvo 2008). In a semi-automatic process, the system finds and suggests elements of a map, and a person manually has to finish the map using the provided information. In the automatic construction process, the user's assistance is not required, and the process creates the map automatically from available resources.

This paper introduces research that addresses the automatic creation of a CM from unstructured text in the morphologically rich Croatian language. It describes the first stage of that research, and presents its threefold direction: a) to gain a better understanding of the problem area, b) to collect information and materials relevant to the research problem, and c) to identify convenient technologies and procedures that can be used in later phases of the research.

This paper is structured as follows. In the second chapter, the CM and CMM-related terms are defined. Literature review of different approaches and previous works to CMM is given in the third chapter. A procedure for the CMM of unstructured textual documents in the Croatian language is proposed and described in the fourth chapter. A short discussion of proposed method is given in the fifth chapter. The sixth chapter provides a brief summary of the paper, and presents a plan for future research activities.

# 2   Concept map mining

This chapter explains and formally defines elements of the CM and the main terms related to a CMM process.

## 2.1   Concept map

A semantic network is a formal structure for representing knowledge as a pattern of interconnected nodes and arcs. Its notation is powerful enough to represent the semantics of natural languages, and can be automatically processed by computer programs (Kramer & Mylopoulos 1987). A CM is a special type of a propositional semantic network that is flexible and oriented to humans. It is designed in the form of a directed graph where nodes represent concepts and arcs represent relationships among them (McNeese et al. 1990).

The educational technique of concept mapping was first attributed to education theorist Novak in 1970s, when his group of researchers described the human learning process as a lifelong process of assimilating new concepts and relations into a personal conceptual framework (Novak & Cañas 2008). Novak adopted the semantic network model and created the CM as a tool for the graphical representation of a learner's conceptual understanding of information in a specific area. His initial idea was that a CM should be drawn free hand by a learner, but only after an initial articulation of major ideas and their classification in hierarchical manner. We now understand, however, that the topology of a CM can take a variety of forms, ranging from hierarchical, to non-hierarchical and data-driven forms.

Formally, a hierarchical CM can be defined (Villalon & Calvo 2008) as a set

$$CM = \{C, R, T\}$$

where

- $C = \{c_0, c_1, ..., c_{n-1}\}$ is a set of concepts . Each concept $c_i \in C; 0 \leq i < n$ is a word or phrase, and is unique in $C$ .

- $R = \{r_0, r_1, ..., r_{m-1}\}$ is a set of relationships among concepts. Each relationship $r_j \in R = (c_p, c_q, l_j), p \neq q; 0 \leq p < n; 0 \leq q < n; 0 \leq j < m$ , connects two concepts, $c_p, c_q \in C$ . Label $l_j$ is a term that labels relationship $r_j$ and represents a conceptual relationship between coupled concepts.

- $T = \{t_0, t_1, ..., t_{s-1}\}; t_{k-1} < t_k < t_{k+1}; 0 < k < s-1$ is a sorted set of hierarchical levels in a CM. Each element $t_k \in T = \{c_0, c_1, ..., c_{r-1}\}; 0 \le r < n$ corresponds to a set of concepts that share the same level of generalization in a CM.

## 2.2 Definition of concept map mining

CMM is a process of extracting information from one or more documents for the automatic creation of a CM. The created map should be a generic summary of a source text (Villalon & Calvo 2008).

From the CMM point of view, a document can be formalized as a set

$$D = \{C_d, R_d\}$$

where

- $C_d = \{c_{d0}, c_{d1}, ..., c_{dn-1}\}$ is a set of all concepts, and
- $R_d = \{r_{d0}, r_{d1}, ..., r_{dm-1}\}$ is a set of all relationships that can be extracted from the document.

Three general phases of the CMM process (Villalon & Calvo 2008) are depicted in the Fig. 1.

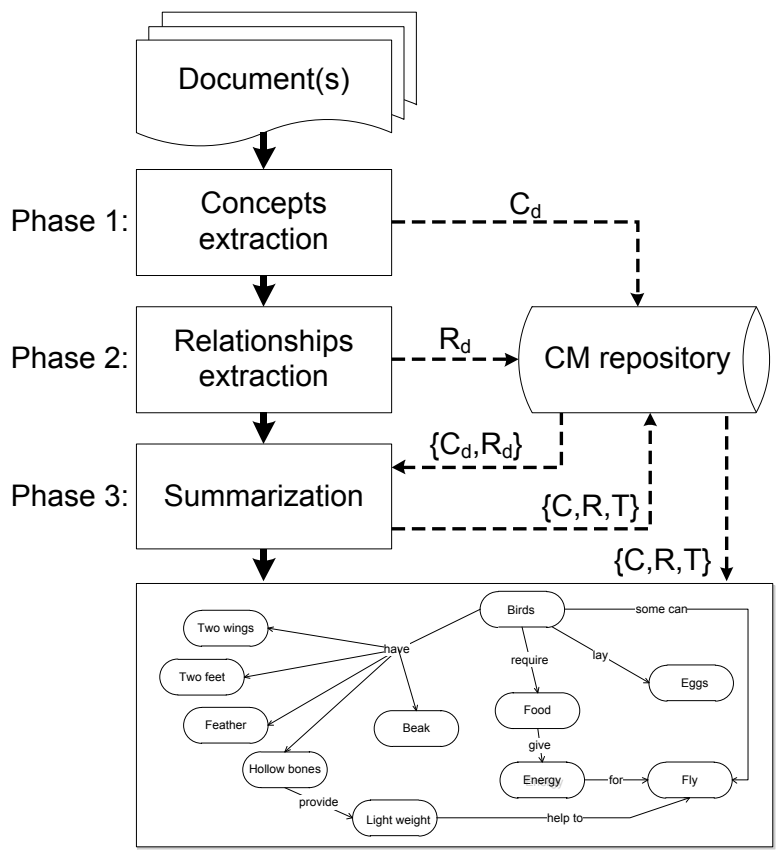

**Fig. 1 General CMM process**

The first phase is the identification and extraction of all members of the set $C_d$. These members are concepts, represented by subjects and objects in the text — usually nouns or noun phrases. For correct and complete understanding of concepts in a map, it is important to consider their connection with synonyms and homonyms. In the natural language processing (NLP) field, that information is usually stored in a repository called a lexicon, where lexical terms are connected with their syntactic or semantic properties. In CMM the same approach can be used, that is, the connection of a concept with its synonyms can be recorder in a simple lexicon. That lexicon can be formalized as a set $Lex = \{lex_0, lex_1, \ldots lex_{n-1}\}$ where each element is a pair $lex_i = (c_i, S_i)$ that contains a concept and a set of concept's synonyms $S_i = \{s_0, s_1, \ldots, s_{m-1}\}$.

A noun can be used as a subject or object in a sentence. When syntactic or semantic dependency between subject and object in a sentence is known, it is possible to extract link that exists between them. Extracting links is a goal of the second phase in this process. A chosen relationship becomes member of the set $R_d$. Each member of that set $r_{dj} \in R_d = (c_{dp}, c_{dq}, l_{dj}), dp \neq dq; 0 \leq dp < dn; 0 \leq dq < dn; 0 \leq dj < dm$ connects two concepts, $c_{dp}, c_{dq} \in C_d$ in a document. Third member of the set $r_{dj}$ is text label $l_{dj}$, which labels a connection between concepts $c_{dp}$ and $c_{dq}$. The final phase of the CMM process is a summarization of the information extracted from document $D$ and the creation of a set $CM = \{C, R, T\}$ that contains concepts, relationships and topological information of the map.

The goal of the CMM process is to produce a CM that is an accurate visual abstract of a source text. That visualization is intended for human analysis, and should not contain too many concepts preferably between 15-25 (Novak & Cañas 2008). In an educational context, the terminology used in a document is important for users, so the CM should be represented using terms that the author uses in the original text.

The source of the CMM technique can be traced back to the early work of Trochim, who proposed a concept mapping process that combines a group activity with statistical analyses (Trochim 1989). During a brainstorming session the group of participants creates a set of statements relevant to the domain of interest. Each participant sorts and rates every statement, creating an individual similarity matrix. All personal matrices are then used to form a group proximity array. The most important statements are chosen using a multidimensional scaling (MDS) and hierarchical cluster analysis. This approach, based on weight calculation and statistical and data mining techniques is still commonly used in many contemporary CMM methods.

# 3  Review of concept map mining approaches

A number of studies have focused on the automatic generation of CMs and similar representations, e.g. topic maps (TMs) from structured and unstructured sources. Some researchers follow the strict definition of a hierarchical CM, while others use knowledge representations that are somewhat more variable. A CMM process is primarily supported by methods used in the NLP field. These methods are briefly described at the beginning of this chapter. In following sub-sections, a short overview of main approaches used in current CMM studies is given.

## 3.1  Methods used in concept map mining

A CMM process can be implemented using NLP methods currently found in tasks supporting information extraction (IE), information retrieval (IR) and automatic summarization. IE is the process of automatically extracting structured information, such as entities and relations, from unstructured textual sources. IR is the area concerned with searching for information in documents and metadata about documents. Lastly, the goal of automatic summarization is to distil content from a source, and present the most important content to the user in a condensed form (Manning & Schütze 1999).

Methods traditionally used in these areas are rule-based statistical and machine learning methods. More recently, there has been interest in combining finite-state machines with conditional-probability models, e.g. maximum entropy Markov models and conditional random fields (CRF) (Manning & Schütze 1999). Most of the classical summarization methods are likewise numerical and based either on a weighting model where the system weights text elements according to simple word or sentence features, or a statistical significance metrics like term frequency-inverse document frequency (TF-IDF).

Machine learning methods often use binary or fuzzy logic to provide accurate extracting based on classification. Such method can be used as a main method, or in a hybrid approach to supply resources to other processes. Contemporary approaches include hybrid techniques using specialized algorithms in combination with third-party party datasets (Das & Martins 2007; Spärck Jones 2007), and summarization based on fuzzy logic and swarm intelligence (Binwahlan et al. 2009).

In the NLP field, numerical methods can be enriched with dictionaries of terms or linguistic tools and techniques (Manning & Schütze 1999). The problem with dictionaries, however, arises from the fact that dictionaries usually come from external resources. It has been previously created for specific language and domain, and therefore requires additional processing to handle the new content and context.

In order to improve processing results, many techniques combine language processing tools, such as tokenizers, stemmers, part-of-speech (POS) taggers, parsers and semantic analysers, with linguistic techniques. A limiting factor for use of such tools and techniques is that many of them are tailored for the English language. This, of course, is problematic when attempting to analyse other languages.

## 3.2   Mining from unstructured text

Unstructured texts in natural languages are commonly used in a CMM process. Considering the number of documents used for one CM creation, there are three groups of applicable techniques. The first group contains techniques that create one CM from a single document (Clariana & Koul 2004; Gaines & Shaw 1994; Kowata et al. 2010; Matykiewicz et al. 2006; Oliveira et al. 2002; Richardson 2007; Saito et al. 2001; Villalon & Calvo 2009; Wang et al. 2008; Willis & Miertschin 2010).  Multiple documents are used as a source in the second group of techniques (Chen et al. 2008; Cooper 2003; Furdík et al. 2008; Hagiwara 1995; Kof et al. 2010; Rajaraman & Tan 2002; Tseng et al. 2010; Valerio & Leake 2006; Zouaq et al. 2011; Zouaq & Nkambou 2008). In the third group, multiple CMs are generated from one long document (Olney et al. 2011). Source documents are usually in shorter forms, such as abstracts or full academic papers, student essays, news articles and medical diagnosis reports. The more lengthy documents used in CMM research include theses and dissertations.

In adaptive learning environments, some researchers analyse students' learning outcomes using CMs automatically created from students' exam results (Bai & Chen 2008; Chen & Bai 2010; Lau et al. 2009; Lee et al. 2009; Sue et al. 2004). Those resources are semi-structured, and embody a conceptual organization that helps facilitate the CMM process.

The goal of most studies in this area is to produce an initial CM model, which can then be used to speed up the process of CM creation for later refinement by a person, or by another automatic process. In this manner, many of created maps are fully completed and contain concepts connected with labelled relationships (Kof et al. 2010; Kowata et al. 2010; Oliveira et al. 2002; Olney et al. 2011; Rajaraman & Tan 2002; Richardson 2007; Saito et al. 2001; Valerio & Leake 2006; Villalon & Calvo 2009; Wang et al. 2008; Willis & Miertschin 2010; Zouaq et al. 2011; Zouaq & Nkambou 2008). In contrast, the goal of one research approach has been to automatically extract terms that are candidates for concepts, followed by a manual construction of the CM from selected concepts (Cañas et al. 2004). Other researchers are in the middle between these two approaches; they automatically create CMs with connected concepts, but without labelled relationships (Bai & Chen 2008; Chen et al. 2008; Chen & Bai 2010; Clariana & Koul 2004; Cooper 2003; Gaines & Shaw 1994; Hagiwara 1995; Lau et al. 2009; Lee et al. 2009; Matykiewicz et al. 2006; Sue et al. 2004; Tseng et al. 2010).

In a process of CMM from unstructured text, authors mostly use methods from NLP field
such as statistical and machine learning methods. Those methods are computationally
efficient, but often not enough precise. In order to improve their precision, authors combine
them with linguistics tools and techniques or databases of key terms important for a target CM
domain. Such hybrid approach helps in more precise extraction of concepts and relationships.

### 3.2.1   The statistical approach

Statistical methods are used for analyse the frequency of terms and their co-occurrence in a
document. They tend to be efficient and transportable, but imprecise because a semantics of
terms are not considered. As they do not depend on a specific language or domain, these
methods can be used in CMM of documents in different languages and different areas. These
methods are commonly used in combination with other methods such as machine learning or
linguistic. The resulting CMs are mainly non-hierarchic, and can be used as scaffolds for the
further exploration of source document content.

Commonly used statistical methods include the analysis of co-occurrences between terms
(Cañas et al. 2004; Clariana & Koul 2004; Cooper 2003; Gaines & Shaw 1994; Tseng et al.
2010), self-organizing maps (Hagiwara 1995) and different term frequency analysing
techniques such as TF-IDF (Lau et al. 2009; Richardson 2007; Valerio & Leake 2006; Zouaq
et al. 2011), latent semantic analysis (LSA) (Villalon & Calvo 2009) and principal component
analysis (PCA) (Chen et al. 2008).

### 3.2.2   Machine learning

Machine learning methods base their functionality on both supervised and unsupervised
learning. The "learned" information is used for extraction of concepts and relationships from
unknown data. Classification, association rules and clustering are the techniques most
commonly used in this process.

The classification technique is used for the extraction of key terms in the TEXCOMON
(Zouaq & Nkambou 2008) software tool. This application uses a simple *Kea* algorithm based
on a naïve Bayes classifier. Lee *et al.* (Lee) uses an *a priori* algorithm and association rules
for the automatic construction of a CM from learners' wrong answers to exam questions. As
the method creates only association rules based on questions that incorrectly answered, it can
miss information associated with correctly answered questions. An improved method (Chen
& Bai 2010) creates association rules for all questions and achieves results that are more
appropriate for use in adaptive learning systems.

A number of methods implement fuzzy reasoning and fuzzy techniques. A hybrid algorithm
called "concept frame graph" (Rajaraman & Tan 2002) uses grammar analysis and fuzzy
clustering techniques for the creation of a CM knowledge base that is represented using concept frames. Each concept frame consists of information about a concept: its name, context, set of synonyms and its relationship to other concepts.

CMM methods that use fuzzy association rules to extract predicates from a learners' historical testing records (Bai & Chen 2008; Sue et al. 2004) are used in adaptive learning environments. In particular, they use a fuzzy association method to search for non-explicit links that exist among concepts. This method employs concept weights combined with fuzzy heuristic. Similar approaches create CMs from newspaper articles (Wang et al. 2008), and messages posted to online discussion forums (Lau et al. 2009).

### 3.2.3 The usage of a dictionary

To define concepts and relationships more precisely, some researchers use ontologies and lists of predefined terms as a seed. Given a single term, it is possible to retrieve terms and relationships that most frequently occur with it across a document collection. Another application of the dictionary is to help minimize the number of terms needed in an extraction phase. With a help of a dictionary, words are grouped by similarity into semantic classes or clusters.

Extracted words stored in a dictionary help in reduce the size of a created CM for the chosen domain. In a simple dictionary a list of only the basic forms of all seed terms is stored. A more complex dictionary contains different forms of terms, meanings and relationships between words and phrases. This type of dictionary can play the role of simple lexicon where terms are connected with their synonyms.

In the biomedical domain, researchers (Cooper 2003; Matykiewicz et al. 2006) used the *Medical Subject Headings* (MeSH) (Lipscomb 2000) and *Unified Medical Language System* (UMLS) (Bodenreider 2004) ontologies as a dictionary. In another research effort (Clariana & Koul 2004), a dictionary was used which consisted of important terms, their synonyms and metonyms selected by an expert. Other examples of dictionaries (Richardson 2007) are those constructed using data from *Computing ontology project* (Cassel et al. 2007), keywords from academic articles (Chen et al. 2008) and index terms from books (Olney et al. 2011; Tseng et al. 2010).

### 3.2.4 Usage of linguistic tools and techniques

Basic statistical and data mining approaches are extensible by using lexical and semantic elements as additional features in calculations. For example numerical technique could make better predictions if that could consider all similar, but slightly different expressions, as a same term. The simplest improvement of numerical techniques is the normalization of inflected words using lemmatization or stemming.

Lemmatization is a process of determining the proper morphological base form (lemma) of a given word through vocabulary and morphological analysis. This process usually involves complex and computationally demanding tasks such as understanding the context and determining the POS of words in a sentence (Manning & Schütze 1999).

Stemming is a computational process that removes the beginnings or endings of words, and thereby identify and remove derivational affixes in systematic way. This approach does not demand a detailed morphological analysis of a text as lemmatization does. Usually, the created root is different from the morphological root of the word. This way of normalization is used when it is sufficient to connect related words to the same stem, even if it is an invalid root (Manning & Schütze 1999).  Stemming is often used for the normalization of texts in less inflected languages, such as English, where it is possible to use simple and efficient stemming algorithms, see for example the *Lovins* (Lovins 1968) or *Porter* (Porter 1980) stemmer.

The problem with linguistic techniques is that linguistic resources must exist for the language been analysed. The majority of linguistic tools and methods used by researchers in CMM are based on the English language (Kof et al. 2010; Lau et al. 2009; Oliveira et al. 2002; Olney et al. 2011; Rajaraman & Tan 2002; Richardson 2007; Valerio & Leake 2006; Villalon & Calvo 2009; Wang et al. 2008; Zouaq et al. 2011; Zouaq & Nkambou 2008). A number of them use *WordNet* (Miller 1995) for lemmatization, POS tagging and terms disambiguation. *WordNet* is designed as a repository of the English language, which groups words into synonym clusters, provides their general definitions, and shows the various semantic relationships among these clusters.

Due to lack of language specific resources, only a few researchers use this approach for mining texts in other languages, e.g., Brazilian Portuguese (Kowata et al. 2010), Chinese (Tseng et al. 2010) , Japanese (Saito et al. 2001) and Slovak (Furdík et al. 2008).

In order to circumvent language-specific problems, Richardson (Richardson) describes a method for the automatic translation of a CM from one language to another. As a practical example, he creates maps from theses and dissertations written in the English language and translates them into Spanish. During a CMM process, syntactic dependencies between noun phrases are recognized using *WordNet's* morphological functions, and are then translated into CM propositions. CMs are summarized using heuristics that employ concepts based on the value of TF-IDF indicator. Created CMs are translated into Spanish using a bilingual dictionary and an algorithm for the translation of individual words and shorter phrases.

A frequent problem that occurs in the recognition of relationships is identifying the noun phrase to which a pronoun phrase refers. This task is known as anaphora resolution (Mitkov 2001). Without proper anaphora resolution, much of the semantic information from a text can be lost, resulting in an incomplete CM where relationships between concepts are missing.

Willis and Miertschin (Willis & Miertschin 2010) use centering resonance analysis for the
creation of CMs reflecting the process of students' assessment. Using linguistics theory, this
method creates a network of nouns and noun phrases to represent main concepts, relationships
and their mutual influence.

## 3.3   Mining from structured textual data sources

Ontologies are used as structured textual data sources in CMM. Their usage in computing has
increased during last decade, especially after introduction of the semantic Web. In the context
of knowledge sharing, an ontology is a description of objects and relationships between them
(Gruber 1993).

CMs and ontologies are quite similar: an ontology can be formalized as a triple *subject-
predicate-object*, and a CM as *concept-relation-concept*. Both of them consist of classes (or
concepts) and relationships among them. Unlike CMs, ontologies are more formal and more
expressive because of their attributes, values and restrictions. The basic approach used for
translating an ontology to a CM is through a direct mapping of ontology classes and
associations into concepts and relationships.

Kim *et al.* (Kim et al. 2006) proposed a way to translate an ontology of the English
vocabulary into a customized CM. The translation process is achieved through a software
agent that directly maps ontological classes and properties to CM propositions. The algorithm
described in (Graudina & Grundspenkis 2008) follows that same approach. The first phase of
that algorithm examines the hierarchy of ontology classes and identifies instances of classes.
Synonyms, intersections and unions among classes are then identifies and translated into
relationships between concepts. Equally, all properties, data types and values become new
concepts. Lastly, the algorithm checks and corrects symmetric and transitive links.

Kumazawa *et al.* (Kumazawa) created a method for visualizing data from a domain-
independent ontology in the form of multiple CMs.  Depending on the user's interests and
perspectives, it enables the visualization of the more domain-dependent parts of a large
ontology. The user chooses a starting concept, and other concepts are then extracted from
ontology according to specific relationships with the starting concept. The depth of extraction
depends on user's choice.

## 3.4   Mining from non-text data sources

Less-commonly used sources for CMM are non-textual entities such as images (Xu et al.
2010), motion stream (Yang et al. 2010), speech (Böhm & Maicher 2006) and video (Badii
et al. 2011).  The first phase of CMM for such sources is the transformation of data into
unstructured text. For audio sources, that transformation process uses automatic speech recognition methods, which are based on hidden Markov models (HMM) (Gales & Young 2007). Elements in visual sources are recognized and transformed using computer vision and pattern recognition methods (Liu et al. 2007; Sebe & Lew 2003). After the text is created, previously described CMM methods for mining from unstructured text are used.

One novel approach for image search (Xu et al. 2010) enables user to enter search terms and their spatial distribution in the form of simple CM. The images in a database are then evaluated using that map. To compare images with a query map, they have to be automatically annotated in the form of a spatial CM where each concept is represented with one salient object in that image. Connections among concepts are represented with their spatial relationships. Salient objects and relationships are found using statistical image segmentation methods such as CRF (Liu et al. 2007).

Yang *et. al.* (Yang) created a compact representation structure of a dance motion in the form of CM which in turn, is used for generation of a dance lessons. After the motion data are captured, they are normalized, and repetitive patterns in a dance motion sequence are identified. A similarity matrix of repetitive patterns is created and all key patterns are found using a classifier. Important relations among those patterns are then identified and used as a basis for the creation of a directed acyclic graph (DAG), which represents sequences of dance moves. The DAG is used for creation of a CM through a process that uses transitive reduction to eliminate redundant relations. In CM, concepts are represented by dance patterns and relations that connect them in a logical order. That map is used for the generation of a dance lessons by extracting patterns in a logical learning sequence and according to the prerequisite relations.

The *SemanticTalk* system observes speech streams and generates a corresponding conceptual graph in the form of a topic map. For speech-to-text conversion, it uses a method based on HMM. Although speech recognition is not error free, the authors argued that for their requirements, the quality of generated word stream is sufficient. Term extraction is performed using a large reference database of about 10 million phrases in specific language. The database is populated by phrases extracted from newspapers, and it can be extended by domain specific terminology (Böhm & Maicher 2006). Nouns that exceed a relevance threshold for a chosen language and domain become key concepts. Relationships are determined based on relations that exist among chosen terms in the underlying reference database. Terms that are not connected with other terms within a defined period are discarded.

DREAM is a semi-automatic framework for multimedia indexing and retrieval. It creates a map of key concepts from multimedia data. Maps created during that process are used as a help in the process of searching of videos (Badii et al. 2011). To build the knowledge base, an automatic labelling module reads the video file and assigns keywords to objects of interest. To find the matching visual objects, the module compare low-level visual features, such as colour, texture, shape, edge, motion activity and trajectory with the visual concepts defined in a vocabulary of objects. For comparison, it uses a term co-occurrences matrix and a k-nearest neighbour algorithm (kNN) (Wu et al. 2007). This is a method for classifying objects based on closest training examples in the feature space algorithm (Badii et al. 2008). After the user examines and updates the list of extracted objects, the resulting text is processed by an engine that uses *WordNet* for POS tagging and semantic normalization. The results are simple sentences where one verb connects two nouns, and complex sentences are built from a few simple sentences. Those sentences are used for creation of a TM.

## 3.5   Evaluation of a CMM process

An important problem emphasized in studies is the need for an objective evaluation method that can be used for the comparison of CMM results. Although most researchers evaluate their CMs, that evaluation is mostly subjective and often based on the personal opinion of only a few experts. If we consider only the quality of extracted concepts, authors argue that their semi-automated extraction methods achieve a concept relevance factor between 65% and 90%. The relevance factor for automatically created maps is reported to be between 50% and 80%.

Humans are good as evaluators of created CMs because they know how to capture important terms and produce a well-formed CM. The problem with human evaluators, however, is that different people look at the source content in their own way, and the process of choosing the key concepts becomes matter of a personal judgment. The quality of evaluation is difficult to control; multiple evaluators will produce somewhat similar, yet distinctively different maps (Dellschaft & Staab 2006). This problem can be reduced by introducing an inter-human agreement among skilful human evaluators who are also familiar with the CMM process and have educational background for creating CMs. Clearly, it is necessary to define their task very precisely (Lin & Hovy 2003). The map agreed under such criteria can be used as a gold standard for evaluation of automatically created CMs.

Villalon and Calvo (Villalon & Calvo 2008) proposed an objective method for evaluating CMs, based on such a gold standard. That method calculates the difference between human-created and machine-generated CMs, using inter-human agreement. According to the authors' recommendation, a gold standard used for evaluation should be a set of CMs created by human annotators who have identified relevant propositions. As a distance measure between the maps, they recommend proven quantitative measures that have been successfully used in different NLP areas such as precision and recall (Manning & Schütze 1999) or Rouge (Lin 2004).

Precision and recall are used for the comparison of referenced concepts with computed concepts and their synonym. This approach is based on the exact matching of strings (Dellschaft & Staab 2006), or alternatively, strings distance metrics (Cohen et al. 2003). In a CMM context, precision can be defined as a proportion between the number of correctly generated concepts and total number of computed concepts; recall is defined as a proportion between the number of correctly generated concepts and the total number of referenced concepts. In order to achieve the best results, it is necessary to have a good balance between the precision and recall values. That balance can be described using the F1-measure, which is the harmonic mean of precision and recall (van Rijsbergen 1997). The CM having the highest F1-measure value is considered to be the best generated map.

A comparison of hierarchical CMMs can also be based on precision and recall measures, but must include information on the topological position of concepts within the map. The procedure for comparison is divided in two phases. In the first, so called local phase, all characteristic subsets of concepts in a CM are identified based on their position and relationship to other concepts in both the calculated and reference hierarchies. Given such subsets, precision and recall can be calculated for them. In the subsequent second phase a global measure is calculated by averaging all results computed during the local phase (Dellschaft & Staab 2006).

# 4   Concept map mining from texts in the Croatian language

This chapter proposes and describes a CMM procedure for automatic creation of CMs from unstructured texts in morphologically rich Croatian language.

## 4.1   Procedure outline

The proposed CMM procedure creates CMs from unstructured texts using statistical and data mining techniques, enriched with linguistic tools. A single CM is created from one document. The first iteration is based on statistical and data mining techniques, with very limited usage of linguistic tools. This approach is suitable for the mining of texts independent of their language origin. Depending on the results, techniques that require deeper linguistic knowledge are used in a second iteration. The main steps of the proposed procedure are shown in the Fig 2.

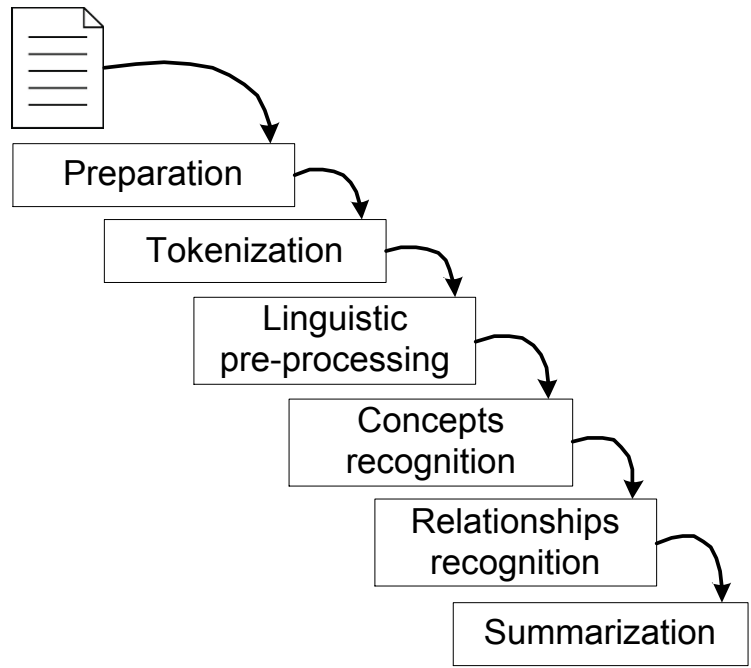

**Fig. 2 General procedure for CMM from unstructured textual sources**

In the first, preparation phase, the text is retrieved from a chosen source. All elements without information are removed and the resulting text, enriched with semantic information, is stored. In the tokenization phase, the retrieved document is divided into sentences, and every sentence into tokens.

The next phase is linguistic pre-processing. From the set of tokens constructed during the tokenization phase, the stop words without information value are removed. In the second iteration, techniques that require deeper linguistic knowledge are used, and every token is associated with its lexical role using the POS tagger. Anaphora resolution is then performed on POS tagged sentences; for problematic pronouns, their corresponding nouns will be determined. Although the tokenization of longer documents can result in a higher dimensionality of the set, this can be reduced by word normalization using lemmatization or stemming. The result of normalization is a reduced set of terms normalized to their basic form.

In the concept recognition phase candidates for concepts are chosen from a set of tokens using the previously created, domain-specific dictionary of key terms. Pre-determined POS tags can assist in this process because nouns and noun phrases are considered as main concepts candidates.

In the relationships recognition phase candidates for relationships are the words semantically connected to extracted concepts. The use of POS tagged verb phrases can help in the extraction of less ambiguous relationships. Combining normalized concepts with links, a set of propositions is created. During the final, summarization phase, the most important propositions are chosen based on their statistical significance; and they serve as a base for CM usage.

## 4.2 Procedure details

### 4.2.1 Data source

This research focuses on the creation of CMs from texts in the Croatian language. Data sources used for analysis are legal documents, and are publicly available from the Web page of the *Official Gazette of the Republic of Croatia* (Narodne novine 2011). It is an accessible source of legal documents written in the correct Croatian language. As legal documents are known to be complex and difficult to understand, CMs can be a helpful addition in supporting and a better understanding of specific documents.

The first step of the proposed process is selection of documents suitable for analysis. Primary choices are papers concerning institutions and groups of people, e.g. laws, regulations, injunctions and rulebooks. Papers focused on individuals in specific positions, such as decisions on appointment or dismissal, are ignored because they are often very short and written using similar phraseology. For these reasons they a not valuable source of information for CMM processing.

The standard semantic structure of a legal text can be useful in a mining process. Important parts of a document that describe semantics include the author, addressee, type and the title of a document, titles of parts, articles and sections in a document, definition of abbreviations and terms, expiry period and connection to other documents. They are collected and used as a document's metadata in later phases of CMM process.

HTML documents are downloaded from the Web page and stored locally. The system will have to determine a correct code page for each document, because documents are encoded using different encoding schemes, e.g. Windows-1250, ISO/IEC 8859-2 and UTF-8. In order to recognize correct code pages, an algorithm checks HTML encoding attribute from the heading of each document. If the value of that attribute does not exist, the algorithm uses heuristics to recognize the correct code page. In addition, it will decode special character entities, like *&* to produce the corresponding set of meaningful characters.

Text and basic semantic information are extracted and analysed; documents that are not suitable for this process are also abandoned. Suitable documents have mark-up and other data with no information value removed from them. Those documents are then stored for further processing. The set of all suitable documents is divided into two parts: a learning set and a test set. The learning set is created by randomly selecting 3/4 of the stored documents; the
remaining 1/4 of the set is used for the test purpose.

### 4.2.2  Creation of a dictionary

A dictionary of the most important terms in the project area is created extracting words and
phrases from the learning document set based on their frequency. Firstly, the stop words are
removed from the set of terms. Stop words are often building elements in multiword phrases,
so this extraction should occur after those phrases are recognized. All terms with a similar
meaning are grouped together using statistical analysis techniques such as LSA or
probabilistic latent semantic analysis (PLSA).

Extracted terms are reduced to their basic form and stored in a hash table. The hash key is the
basic form of a term. As a part of the dictionary, an acronym-mapping table is created to
connect acronyms to mapping phrases. Abbreviations depend on a specific document and are
not stored in the dictionary. Nonetheless, a list of phrases connected with each abbreviation is
created, and a probability of connection is determined. Those probabilities can be used in later
phases for resolving ambiguities associated with an abbreviation's meaning.

### 4.2.3  Tokenization

During the tokenization phase, basic language elements are identified. These elements are
usually words or phrases separated by non-alphanumeric tokens such as white spaces and
punctuation marks. Initially a simple strategy is used to split sentences based on the presence
of non-alphanumeric characters.

All words are converted to lowercase. There are exceptions, especially for abbreviations,
hyphenations, digits and some name entities. An abbreviation-mapping table is created to map
all abbreviations in a document to their mapping phrases. Many proper nouns are derived
from common nouns and distinguished only by case. To recognize them, a simple heuristic
algorithm is used that leaves capitalized mid-sentence words as is, and converts to lowercase
the capitalized words at the beginning of sentence.

In creating a token list, it is important to record the connection of each word with its sentence
because later feature analyses will use this information. A weight is assigned to every term;
terms extracted from a document's semantic data have a larger weight than terms from
ordinary text. That weight is used in the process of summarizing a CM. This stems from our
presumption that semantic data carry more precise information than ordinary text.

### 4.2.4  Linguistic pre-processing

During the linguistic pre-processing phase source data are transformed into the appropriate form. This phase consists of several language-dependent NLP steps that provide annotations to text resources.

At the beginning of this phase, stop words are removed from the set of tokens and all words are normalized. In highly inflected languages such as Croatian, this phase is very important for reducing inflected words to their common base form. Word normalization can be achieved using lemmatization or stemming.

Several studies have examined different methods of stemming (Kalpic 1994; Lauc et al. 1998; Ljubešic et al. 2007; Snajder et al. 2008; Šnajder & Dalbelo Bašic 2009) and lemmatization (Lujo 2010; Tadic & Fulgosi 2003; Šnajder 2010) in the Croatian language. Generally, in morphologically rich languages lemmatization techniques achieve much better results than stemming. A problem with the practical usage of lemmatization is that it is more computationally demanding, and therefore, not always suitable for many applications.

In the first iteration of this phase, stemming is used as a normalization technique. A simple stemmer for the Croatian language is built using a rule-based approach similar to those described in (Ljubešic et al. 2007). In the second iteration, depending on the achieved results, techniques that require deeper linguistic knowledge will be used. In that iteration, lemmatization will be used for the normalization. During the normalization activity extracted words will also be POS tagged.

Initial analysis of sample text shows that in this process, anaphora would be a minor problem. This reasoning stems from the fact that authors of legal documents try to be as precise as possible in their writing style and generally avoid the use of pronouns.

### 4.2.5 Concepts extraction

Concepts extraction is a process of discovering potential candidates for concepts in a CM. Generally, the subject of a sentence represents the first concept, and the second concept is represented by the object of a sentence.

The algorithm for extraction that is used in this phase is based on the previously created dictionary. A set of rules is created and all terms whose base forms are found in the dictionary are directly marked as concept candidates. In addition, all adjacent words that frequently occur in the neighbourhood are marked as concept candidates, and a connection among them is established. A word is considered as neighbour if it is part of the same sentence as a concept, and is within a specific distance from that concept. TF-IDF indices are used for a frequency calculation. In the second iteration of this research, knowledge of a POS role for the chosen term can be a useful addition to this process, as nouns and noun phrases are main candidates for concepts.

### 4.2.6 Relationships extraction

Concepts in a CM are semantically related and, during this phase, links between them is established. The type and label of a relationship between two concepts in a simple sentence can be identified by the main verb in the sentence. For each pair of concept candidates, all words positioned in their neighbourhood are temporarily saved as candidates for a link. A set of rules is created for the extraction of link candidates from that temporary storage set. Extraction rules are based on the frequency of their appearance in the concepts' neighbourhood.

If the frequency of appearance of two concepts in the same sentence is high, then the degree of the relationship between them is high. If the frequency of appearance of two concepts in the neighbouring sentences is high, then a degree of relationship between them is medium. Frequency is calculated using the value of TF-IDF indices. Each relation having a high or medium degree is saved as a three-element set: the two adjacent concepts, label of the link between them and the strength of that relation. Each set represents one proposition, and is used as input in the succeeding summarization phase. In the second iteration of this research, POS tags can be used since verbs and adverbs are main candidates for relationships.

### 4.2.7 Concept map summarization and generation

The result of the summarization phase is the CM that provides an overview of the document's contents with minimal redundancy. In the first iteration, statistical techniques used for summarization of dictionary terms, such as LSA or PLSA, are used on propositions created in the previous phase. The relative importance of propositions within a document is calculated, and top propositions are extracted. Propositions with the higher calculated values are positioned higher in the CM hierarchy; the strongest proposition is marked as the starting one. Based on an examination of summarization results in the first iteration of research, specific technique will be chosen for use in the second iteration.

An optimal number of concepts in the final CM is determined, and then summarized propositions are stored in CXL format (Cañas et al. 2006). Maps in that format are XML-based and can be visualized using different concept mapping tools.

### 4.2.8 Evaluation of created concept maps

An evaluation of automatically created CMs is performed using the gold standard CMs created by human annotators. Human annotators create CMs from randomly chosen documents in a test set. In order to get statistically significant results, each map is created by several evaluators. Propositions extracted from those maps provide a foundation for measuring the agreement among human annotators. Inter-human agreement among created documents is measured using kappa statistics or similar methods that measure the extent to
which annotators agree in their interpretations (Gwet 2010).

Automatically created CMs are compared with a gold standard map of the same document
using precision and recall, or Rouge measures commonly used for evaluation in the NLP area.
The calculated difference between the gold standard and generated maps is used to tune the
CMM process.

# 5   Discussion

Most of the previous CMM methods have been applied and customized for small document
collections in a known domain. These techniques are mainly used to evolve a language model,
or a set of rules, from training documents, and then apply that model or rules to new texts. In
such a closed environment approaches that use profound linguistic knowledge and techniques
can achieve good results. Models determined in this manner can be used effectively on
documents that are similar to the training documents, but behave poorly when applied to
documents of differing similarity. As a result, this approach cannot be efficiently used in an
open environment like the Web due to the diversity of problem domains, high usage of
computer resources and the cost of creating an equally diverse set of training documents. For
use in such environments, researchers should abandon strategies that require extensive
domain and linguistic knowledge, and adopt knowledge-poorer strategies.

The problem with automatically created CMs is that the number of extracted concepts is often
huge or too small, and often contains concepts irrelevant to the problem domain. It is difficult
to find representative sets of concepts, correct complex phrases, labels and directions of links.

Many similar CMM research approaches that use linguistic tools and techniques were dealing
with low or moderately inflective languages such as English, Portuguese, Japanese or
Chinese. The method described in this paper is focused on CMM from texts in the Croatian
language, which is a highly inflective language. With smaller amendments that method can
also be used for CMMs from text in other morphologically rich languages. Morphologically-
rich languages characterized by productive morphological processes such as inflection can
produce a very large number of word forms for a given root. The increased size of a data set
causes additional problems in data processing, especially during creation of a dictionary.
Also, it can place heavy demands on computational resources.

The data source for this research consists primarily of legal texts in the area of science and
education. These documents vary in size, from very short to very long. Other studies are
focused on mining smaller text sets of similar sizes such as essays, newspaper articles or
academic papers, whereas some studies focus on longer text sets such as theses and dissertations. When longer documents are used as data source, it is easier to extract a complete set of concepts, but data processing demands for such documents can be more extensive.

The CMM method described in this manuscript uses unstructured text and semantic information embedded in a document. Semantic information extracted from that data will have a larger weight than that extracted from ordinary text. To the best of our knowledge, this approach, which is common for summarization techniques, has not been used as a part of CMM methods described in other studies.

# 6  Conclusion

A fully automatic production of a human-quality CM from a given document is difficult and has not yet been satisfactorily resolved. It is not enough just to extract words from a document; one must also identify and label relevant concepts and relationships among them.

This paper deals with CMM methods and approaches. In the initial section, a definition and a literature review of that area are given. In the following chapter, a procedure for producing CMs from unstructured texts written in the Croatian language is proposed and described in detail.

The next step of this research is the practical implementation of the proposed procedure. Research is conducted in two iterations. In the first iteration, a "linguistically poor" approach using statistical and data mining methods is implementing. For the creation of a prototype in that phase, existing programming tools and applications are used as much as possible. At the end of that iteration, a test of the prototype and an evaluation of created maps will be conducted. Based on the results, mining and evaluation procedures will be adjusted.

One of goals that we shall try to achieve in this research is to create a system that is applicable as a real-time operation. This means that precision may be sacrificed for a better and faster response time. Considering this, we shall have to make a decision about the use of richer linguistic tools and techniques that are to be used in later stages of the research. If tools chosen for a specific task cannot fully achieve expected results, we shall modify or rebuild them.

In the second iteration of this research, a modified procedure and set of tools will be used for the creation and evaluation of a second version of the prototype. In the final part of the research, created features will be integrated in the personal learning environment application.